\documentstyle[times,epsf]{mn}

\title[3C~220.1's X-ray core and cluster]{{\it Chandra\/}
measurements of 3C~220.1's X-ray core and cluster}

\author[D.M. Worrall et al.]{D.M.~Worrall,$^1$ M.~Birkinshaw,$^1$ 
M.J.~Hardcastle$^1$ and C.R.~Lawrence$^2$\\
$^1$Department of Physics, University of Bristol, Tyndall Avenue,
Bristol BS8 1TL, UK\\
$^2$Jet Propulsion Laboratory 169-506, Pasadena, CA 91109, USA}

\begin{document}

\maketitle

\label{firstpage}

\begin{abstract}

We report results of an 18~ks exposure with the ACIS instrument on
{\it Chandra\/} of the powerful $z = 0.62$ radio galaxy 3C~220.1.  The
X-ray emission separates into cluster gas of emission-weighted $kT$
$\sim 5$ keV and 0.7-12~keV luminosity (to a radius of 45 arcsec) $5.6
\times 10^{44}$ ergs s$^{-1}$, and unresolved emission (coincident
with the radio core).  
While the extended X-ray emission is clearly thermal in nature, a
straightforward cooling-flow model, even in conjunction with a
point-source component, is a poor fit to the radial profile of the
X-ray emission. This is despite the fact that
the measured properties of the gas
suggest a massive cooling flow of $\sim 130$ M$_\odot$ yr$^{-1}$, and
the data show weak evidence for a temperature gradient.
The central unresolved X-ray emission has a power-law
spectral energy index $\alpha \sim 0.7$ and 0.7-12~keV luminosity
$10^{45}$ ergs s$^{-1}$, and any intrinsic absorption is relatively
small.  The two-point spectrum of the core emission between radio and
X-ray energies has $\alpha_{\rm rx}$ = 0.75.  Since this is a flatter
spectrum than seen in other sources where the X-ray emission is
presumed to be radio-related, regions close to the AGN in this source
may dominate the central X-ray output, as is believed to be the case
for lobe-dominated quasars.  Simple unification models would be
challenged if this were found to be the case for a large fraction of
high-power radio galaxies.

\end{abstract}

\begin{keywords}
galaxies:active -- 
galaxies:clusters:individual: 3C~220.1 --
galaxies:individual: 3C~220.1 --
radiation mechanisms: non-thermal --
X-rays:galaxies
\end{keywords}

\section{Introduction}

Powerful radio sources of linear size $> 50$ kpc, through their very
existence, point to gaseous atmospheres on scales of at least the
radio-source diameter.  Indeed, evidence suggests that many, and maybe
most, such sources have lobe minimum pressures which lie below the
external thermal pressure (Hardcastle \& Worrall 2000).  High-redshift
sources are thus useful tracers of galaxy groups and clusters for
cosmological studies, and, whereas X-ray selected clusters are biased
towards only the more luminous clusters at higher redshift, radio
selection should provide a more representative sample of cluster
X-ray properties for tests of structure-formation theories and
cosmological parameters.

{\it ROSAT\/} pioneered the detection of radio-source atmospheres at
high redshift.  The best catalogue available for the selection of the
most luminous radio galaxies is 3CRR (Laing, Riley \& Longair 1983),
which is complete to 10.9~Jy at 178~MHz ($\delta > 10^\circ$ and $|b|
>10^\circ$).  There are 38 powerful radio galaxies with linear size $>
50$ kpc at $z > 0.6$.  Of these, 12 were observed in {\it ROSAT\/}
pointed observations, and most were detected (Table~\ref{cluster}).
The most significant detections (3C~220.1, 280, 324, and 294) provided
evidence for source extent.  The X-ray luminosities in
Table~\ref{cluster} assume all the X-rays originate in plasma with
0.3~solar metallicity and 5~keV temperature, except for 3C~220.1 and
280 where a modelled component of compact emission has been
subtracted.  The total extended luminosity may exceed quoted values
since the on-source extraction radii were relatively small (to maximize
detection significance).

\begin{table}
\caption{{\it ROSAT\/} pointings at $z > 0.6$ 3CRR galaxies}
\label{cluster}
\begin{tabular}{llllc}
Name &
$z$ &
inst. &
ref &
$L_{2-10~\rm keV}/10^{44}$
\\
&&&& ergs s$^{-1}$
\\
3C 220.1 & 0.61 & HRI & 1,3 & 4.2\\
3C 220.3 & 0.685 & PSPC & 2,3,5 & $<$ 1.7\\
3C 247 & 0.7489 & HRI & 3 & $<$ 0.47\\
3C 277.2 & 0.766 & PSPC & 3,5 & 0.5\\
3C 263.1 & 0.824 & HRI & 3,5 & 8.2\\
3C 289 & 0.9674 & HRI, PSPC & 3,4,5 & 1.2 \\
3C 280 & 0.996 & PSPC, HRI & 2,3,5 & 1.6 \\
3C 356 & 1.079 & HRI, PSPC & 3,5 & 1.5 \\
3C 368 & 1.132 & PSPC  & 3,4 & 2.1 \\
3C 324 & 1.2063 & HRI, PSPC  & 3,5 & 4.2 \\
3C 13 & 1.351 & PSPC  & 3,5 & $<$ 4.2\\
3C 294 & 1.781 & HRI & 3,5 & 6.4\\
\end{tabular}
\medskip
\begin{minipage}{\linewidth}
Refs: 1. Hardcastle et al. (1998); 2. Worrall et al. (1994); 3.
Hardcastle \& Worrall (1999); 4. Crawford \& Fabian (1995); 
5. Crawford \& Fabian (1996).  Ref.~5 contains additional 3C
sources not in the complete 3CRR subset
\end{minipage}
\end{table}

\begin{table*}
\caption{3C~220.1 observed with {\it Chandra}}
\label{xrayobs}
\begin{tabular}{lccllllc}
$z$ &
kpc/ &
Galactic N$_{\rm{H}}$ &
\multicolumn{2}{c}{J2000 X-ray core position} &
shift &
Date  & 
Screened 
\\
& arcsec & (cm$^{-2}$) & \multicolumn{2}{c}{J2000 radio core position} &  (arcsec) 
& & Exposure (ks) \\
0.620 & 9.0 & $1.93  \times 10^{20}$ & 09 32 39.84 & 
$+$79 06 31.9 & 0.66 & 1999 Dec 29/30 & 18.121 \\
&&& 09 32 39.646 $\pm 0.004$ & $+$79 06 31.53 $\pm 0.02$ &&& \\
\end{tabular}
\medskip
\begin{minipage}{\linewidth}
We use $H_o = 50$ km s$^{-1}$ Mpc$^{-1}$, $q_o = 0$, throughout.
Radio position (quoted with errors) is based on data shown in
Fig.~\ref{profile}.
\end{minipage}
\end{table*}

\begin{table*}
\caption{PRF parameter values in function
$(A_1 + B_1 r + C_1 r^2)e^{-{r^2 \over 2 S_1^2}}+
(A_2 + B_2 r + C_2 r^2)e^{-{r^2 \over 2 S_2^2}}+
(A_3 + B_3 r + C_3 r^2)e^{-{r^2 \over 2 S_3^2}}$
}
\label{prfpars}
\begin{tabular}{rrrrrrrrrrrr}
$A_1$ &
$B_1$ &
$C_1$ &
$S_1$ &
$A_2/10^{-2}$ &
$B_2/10^{-2}$ &
$C_2/10^{-3}$ &
$S_2$ &
$A_3/10^{-5}$ &
$B_3/10^{-6}$ &
$C_3/10^{-7}$ &
$S_3$
\\
1.874 &
$-$2.995 &
1.395 &
0.4494 &
0.9572 &
$-$0.4931 &
0.8257 &
1.665 &
5.773 &
$-$5.193 &
1.362 &
11.42
\\
\end{tabular}
\medskip
\begin{minipage}{\linewidth}
$r$ is in arcsec.
The profile is normalized to $\sim$~1.03 rather than 1.0; this is
correct for use with 0.1~arcsec bins and any radial-profile 
extraction routine which works like {\sc iraf/imcnts}
in considering square pixels as falling
entirely in or out of an annulus rather than weighted by area.
See Worrall et al.~(2001) for more details.
\end{minipage}
\end{table*}

Under unification models, radio-loud quasars are believed to be
powerful\footnote{isotropic luminosity at 1.4~GHz $> 10^{27}$ W
Hz$^{-1}$} radio galaxies whose relativistic jets are at a small angle
to the line of sight, and so should be associated with similar X-ray
atmospheres.  Preliminary confirmation comes from the {\it ROSAT\/}
detection of extended X-ray emission around four X-ray-bright 3CRR
quasars at $z > 0.4$ (Crawford et al. 1999; Hardcastle \& Worrall
1999), one of which (3C~254) is at $z > 0.6$.

The atmospheres are not the only components of X-ray emission in
powerful radio sources.  Core-dominated quasars, where the dominant
core X-rays are undoubtedly radio related and highly beamed, have a
small dispersion of radio to soft X-ray two-point spectral index,
$\alpha_{\rm rx}$, centered around a value of $\sim 0.85$ (Worrall et
al.~1994; Worrall 1997).  In broad emission-line objects, such as
lobe-dominated quasars, a smaller $\alpha_{\rm rx}$ is normally seen.
This is thought to be due to the fact that in such objects there is
little or no steep-spectrum ($\alpha_{\rm rx} \approx 0.85$) beamed
emission to dominate the flatter spectrum ($\alpha_{\rm rx} < 0.85$)
emission arising from regions close to the AGN.  In powerful
narrow-line radio galaxies, X-rays from the regions close to the AGN
are expected from unification models to suffer high absorption from
the torus of gas and dust invoked to obscure the broad emission-line
regions, and so any flat-spectrum ($\alpha_{\rm rx} < 0.85$) component
of the emission should be weak.  The nearby powerful radio galaxy
Cygnus~A supports this picture, in that a component of highly
absorbed ($N_{\rm H} \sim 4 \times 10^{23}$ cm$^{-2}$) emission was seen at
hard X-ray energies with {\it EXOSAT\/} and {\it Ginga\/} (Arnaud et
al.~1987; Ueno et al.~1994) while a significant fraction of the soft
core X-ray emission seen with {\it ROSAT\/} could plausibly be radio
related with $\alpha_{\rm rx} \sim 0.85$ (Worrall~1997).  For the
powerful radio galaxy 3C~280, where point-like and extended components
in the {\it ROSAT\/} data were separated, $\alpha_{\rm rx}$ was
measured to be 0.85 (within errors), suggesting it was predominantly
radio-jet-related soft X-ray emission which was being detected
(Worrall et al.~1994).

Sensitive, high-spatial resolution observations now possible with the
{\it Chandra X-ray Observatory\/} (Weisskopf et al.~2000) can probe
both the extended and compact emission from high-power radio galaxies,
using them as signposts to clusters with significant atmospheres and
as a test of unification models.  In this paper we report results for
the first high-power radio galaxy we have observed with {\it
Chandra\/}.  3C~220.1 was selected for study since it was the {\it
ROSAT\/}-observed $z > 0.6$ 3CRR radio galaxy with the highest count
rate, and the only source other than 3C~280 for which it had been
possible to attempt X-ray component separation with the {\it ROSAT\/}
data (Hardcastle, Lawrence \& Worrall~1998).

\section{{\it Chandra\/} Observations}

\begin{table}
\caption{Energy weighting for the PRF, 0.3 - 8 keV}
\label{prfweights}
\begin{tabular}{ll}
Energy (keV) &
Weight \\
0.65 & 0.474 \\
1.35 & 0.343 \\
2.4 & 0.081 \\
3.4 & 0.057 \\
4.5 & 0.021 \\
5.5 & 0.024 \\
\end{tabular}
\end{table}

We observed 3C~220.1 with the Advanced CCD Imaging Spectrometer (ACIS)
on board {\it Chandra\/} on 1999 December 29/30.  The target was near
the aim point on the back-illuminated CCD chip S3.  The observation
was made in full-frame mode, and chips S2, S4, I2 and I3 were also
active.  The data provided to us had been processed using version
R4CU4UPD5 of the pipeline software, with subsequent custom processing
to alleviate the effect of a large (about 8 arcsec) positional offset
which we reported present in the data (and which was subsequently
found to affect all data processed during a roughly 7 week period).
We followed the ``science threads'' from the {\it Chandra X-ray
Center\/} (CXC) for {\sc CIAO v 1.1.5} to make the recommended
corrections to these data, and in particular to apply the appropriate
gain file, acisD1999-09-16gainN0004.fits.  After screening out about
10 per cent of the observation to avoid intervals of high background,
the exposure time was 18.121~ks.  The radio core position is known
with high precision (Table~\ref{xrayobs}).  The separation of the
X-ray core position and the radio position is 0.66 arcsec, within the
astrometric accuracy of {\it Chandra\/} (see
http://asc.harvard.edu/mta/ASPECT/).

\begin{figure*}
\epsfxsize 14.0cm
\epsfbox{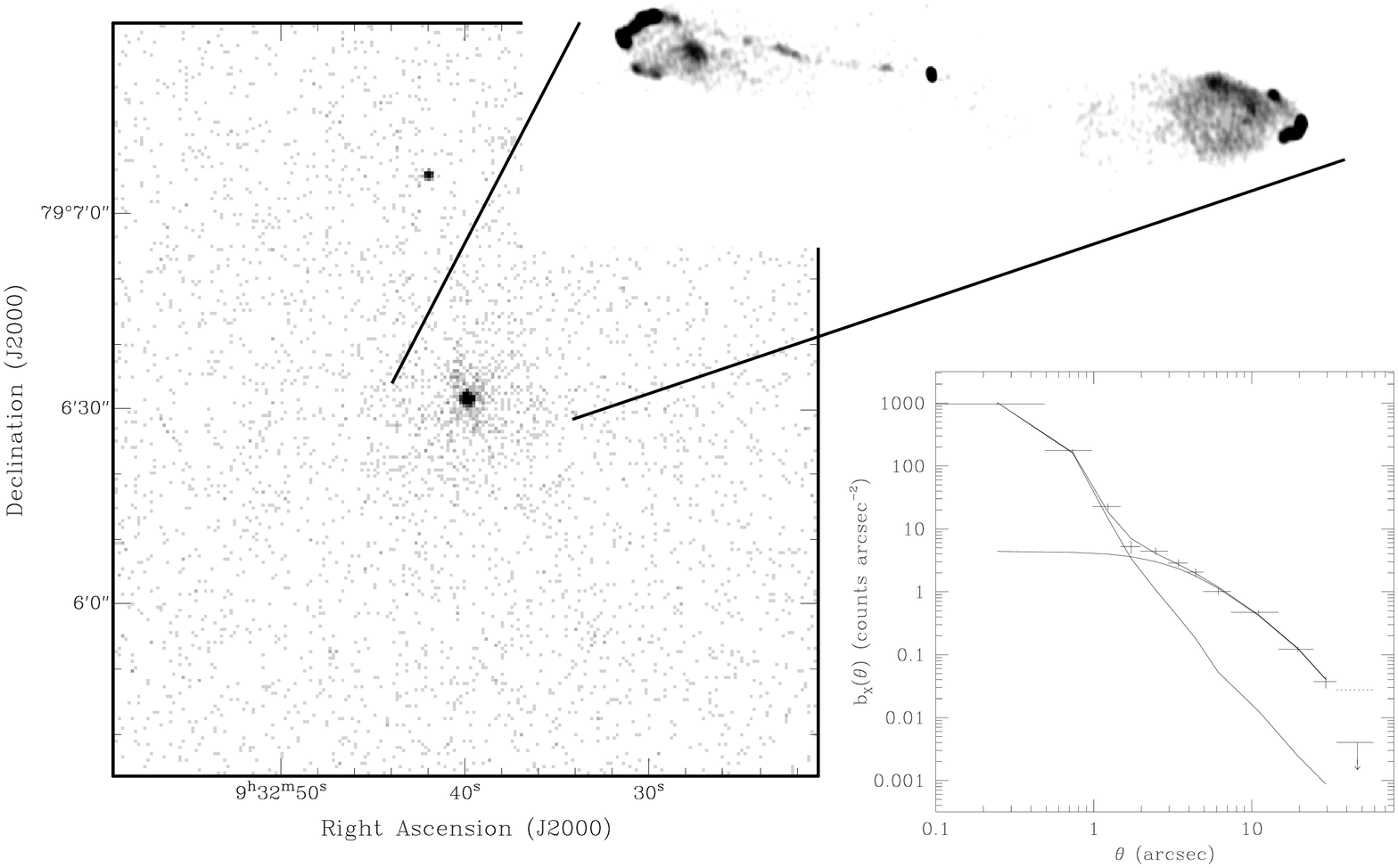}
\caption{
Main picture on left is the unsmoothed {\it Chandra\/} image with 0.5
arcsec pixels centered on 3C~220.1.  An unrelated X-ray point source
lies to the north.  The insert shows an 8.4~GHz VLA A and B array
radio image of 3C~220.1 with resolution of 0.3 $\times$ 0.2 arcsec.
On the right is the background-subtracted radial profile of the X-ray
data for 3C~220.1, plotted as counts per square arcsec,  $b_x$, as
a function of angular distance from the core.  Solid curves show the
two components of the best-fit model (point-like emission and a
$\beta$-model with $\beta=0.5$, $\theta_{\rm cx}$ = 3.55 arcsec;
$\chi^2 = 12$ for 8 degrees of freedom) plus their sum.  The
contribution of the model to the background region (here an annulus of
radii 35 and 60 arcsec), taken into account in the fitting, is shown
dotted.}
\label{profile}
\end{figure*}

We followed the procedure described in Worrall, Birkinshaw \&
Hardcastle (2001) to find an analytical description of the Point
Response Function (PRF) appropriate to our observation of 3C~220.1.
The CXC-released PRF library was used to create an image of the PRF
appropriate to the chip position and energy weighting of counts from
the central region of 3C~220.1.  The same analytical function as used
in Worrall et al.~(2001) was found to give a good fit to the radial
profile extracted from the PRF image, and the fitted parameter values
are given in Table~\ref{prfpars}.  The energy weightings are in
Table~\ref{prfweights}.
The resulting profile has a half power
diameter (HPD) of 0.8 arcsec, and full width half maximum (FWHM) of
0.58 arcsec.  A slightly greater fraction of the PRF is in the
large-scale wings than for the low-power radio galaxies from the B2
sample for which PRF parameters are given in Worrall et al. (2001);
this is due to 3C~220.1's harder X-ray spectrum.

\section{X-rays from the cluster and core}

\begin{figure}
\epsfxsize 7.0cm
\epsfbox{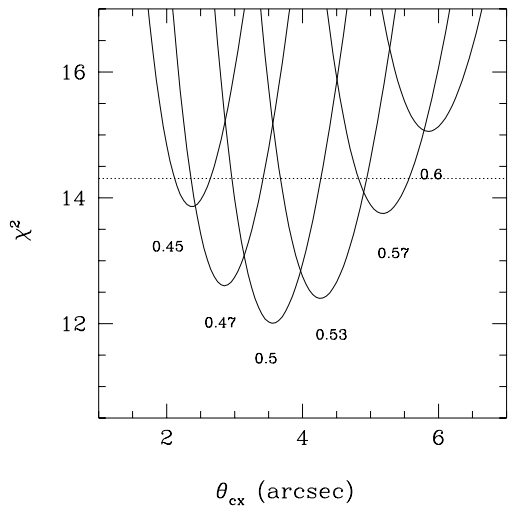}
\caption{
Results of fitting a $\beta$-model plus point source to the {\it
Chandra\/} radial profile of Fig.~\ref{profile}. 
$\chi^2$ versus core radius of
the $\beta$ model is shown, for representative values of $\beta$;
these two parameters are highly correlated.
The dotted line is at $\chi^2_{\rm min} + 2.3$, corresponding to
$1\sigma$ for 2 interesting parameters. Results in \protect{Table~\ref{xraytab}}
include a small additional error arising from varying the choice of
background annulus.
}
\label{chi}
\end{figure}

The inference from {\it ROSAT\/} of both point-like and extended X-ray
emission in 3C~220.1 (Hardcastle et al.~1998) is verified in a
dramatic way with {\it Chandra\/}, visible by eye in the image and
confirmed by detailed analysis of the radial profile
(Fig.~\ref{profile}).  The radial profile is centrally spiked, fitting
well the PRF in the inner bins, and extended emission is detected out
to a radius of $\sim 45$~arcsec. 
The extended emission contributes some noticeable excess
in the reference region ($35''-60''$) which we have used as background for
our profile, and this excess is taken into account in the model fitting.
As in our {\it ROSAT\/} analysis, we have
modelled the radial profile as a composite of the PRF and a $\beta$
model, the form representing gas in hydrostatic equilibrium.  The fit
is excellent.  Fig.~\ref{chi} illustrates acceptable values of $\beta$
and core radius, $\theta_{\rm cx}$, parameters which were poorly
constrained with {\it ROSAT\/}.  Values and errors for a large set of
parameters are listed in Table~\ref{xraytab}, using the procedure
described by Worrall \& Birkinshaw (2001) to ensure realistic
uncertainties.  Our conclusion from {\it ROSAT\/} that the minimum
pressures in the radio lobes lie below the external thermal pressure
(Hardcastle \& Worrall~2000) is confirmed by the {\it Chandra\/} data,
and a similar conclusion can be drawn for the 9~arcsec-long eastern jet
(see Tables~\ref{xraytab} and \ref{radiotab}).

\begin{figure}
\epsfxsize 5.0cm
\epsfbox{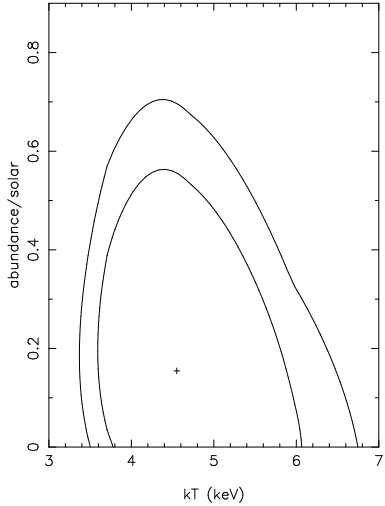}
\caption{
The background-subtracted X-ray counts within an annulus of radii
2 and 45 arcsec
are heavily dominated by the resolved component of emission
\protect{(Fig.~\ref{profile})}. 
Their spectrum fits a Raymond-Smith thermal model with Galactic
absorption
\protect{(Table~\ref{xrayobs})}, with $\chi^2 = 41$ for 41 degrees of freedom.
Uncertainty contours for
temperature and abundance are shown, corresponding
to 90 per cent confidence levels for one (inner contour) and two (outer
contour) interesting parameters.
}
\label{gas}
\end{figure}

Many of the source parameters rely not only on the spatial modelling
of the components, but also their spectra.  Since the unresolved
component is so strong, a radius of 2~arcsec marks a reasonable
boundary between regions dominated by the central component and those
dominated by the extended emission. The extended emission gives a good
fit to thermal emission seen through Galactic absorption
(Fig.~\ref{gas}), and no significant improvement in the fit
($\Delta \chi^2 < 1.2$)
is obtained if the absorption is allowed to be a free parameter.  The
X-ray spectrum alone does not require a thermal origin: no obvious
X-ray line features are seen, and the spectrum can be fit with a power
law of photon index ($\alpha + 1$) $\sim 2.3$ and an absorption of
$N_H \sim 10^{21}$~cm$^{-2}$, in excess of the Galactic value
(Table~2).  However, the radial symmetry of the emission rules out a
non-thermal origin related to the relativistic particles in the radio
jets and lobes, and we conclude that the emission is undoubtedly
predominantly thermal in origin.  Independent support for the presence
of a galaxy cluster arises from the luminous gravitational lens arc
originally detected in an optical ground-based image (Dickinson~1994)
and now known from HST data to have a redshift of $z = 1.49$
(M. Dickinson, private communication, 1997).

\begin{figure}
\epsfxsize 5.0cm
\epsfbox{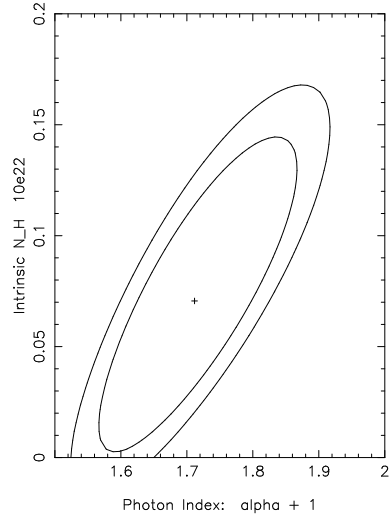}
\caption{
The background-subtracted X-ray counts within a radius of 2 arcsec
are heavily dominated by the unresolved component of emission
\protect{(Fig.~\ref{profile})}.  Their spectrum
fits a power law with a small amount of
intrinsic absorption in addition to that from our Galaxy
(Table~\ref{xrayobs}), with $\chi^2 = 44$ for 42 degrees of freedom.
Confidence levels of contours are as in Fig.~\ref{gas}.
}
\label{core}
\end{figure}

Any attempt to fit the central emission to a single-temperature
thermal model finds abundances of zero and a very hot temperature of
$kT$ greater than about 9~keV.  Indeed, the spectrum is harder than
that of the diffuse X-ray emission (as verified by a colour image of
the field), and fits well a power law with modest intrinsic absorption
(Fig.~\ref{core}).  Ota et al.~(2000) measured the total emission
within a radius of 3~arcmin in a 40~ks ASCA exposure to have a
temperature of $kT = 5.6 (+1.5, -1.1)$~keV (90 per cent confidence for
one interesting parameter), which is in good agreement with our
results (Fig.~\ref{gas}) despite the non-negligible contamination of
the central point-source emission in the ASCA spectrum: the point
source provides 46 per cent of the 0.3-8 keV {\it Chandra\/}
counts over the same region.

\begin{figure}
\epsfxsize 5.0cm
\epsfbox{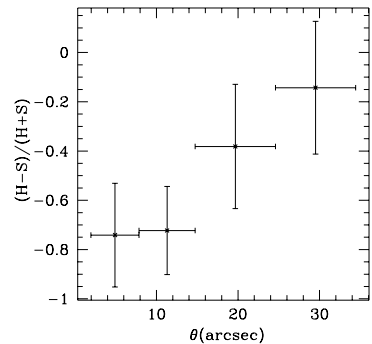}
\caption{ Hardness ratio as a function of radius for the extended
excess, corrected for background and the wings of the point source
component (see Fig.~\ref{profile}).  H = hard (3--8~keV) counts; S =
soft (1.5--3~keV) counts. At low significance there is a trend for the
extended emission to decrease in hardness towards the centre,
indicative of cooling gas.  }
\label{hardness}
\end{figure}

\begin{figure}
\epsfxsize 7.4cm
\epsfbox{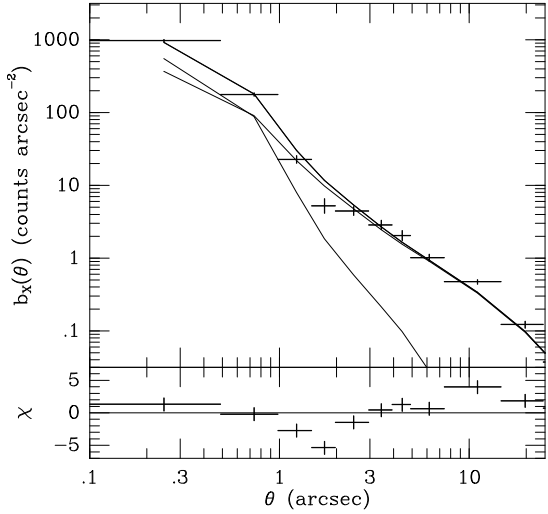}
\caption{ A straightforward cooling-flow model with density $\propto
(\theta/\theta_{\rm cool})^{-1.5}$ and temperature $\propto
(\theta/\theta_{\rm cool})^{0.5}$ within a cooling radius of
$\sqrt{2}~\theta_{\rm cx}$ (Hardcastle et al.~1999), fitted in
combination with a point-source, is too spiked for the radial
profile of the counts.  
Solid curves show the modelled
point-like emission (narrow curve), cooling-flow model (broader curve) 
and their sum.
In this example the $\beta$ model (at $\theta
> \theta_{\rm cool}$) has $\beta = 0.5$ and $\theta_{\rm cx} =
3\farcs55$ (Table~\ref{xraytab}), giving $\chi^2 = 61$.  }
\label{cooling}
\end{figure}

There is some evidence that the temperature of the gas increases to
larger radius, as shown in Fig.~\ref{hardness}. Taking counts only in
an annulus of radii 10 and 45 arcsec, we find an emission-weighted
$kT$ of 8.5 keV (with a 90 per cent uncertainty range for one
interesting parameter of 6.2 to 12.2 keV) which is hotter than the
overall emission-weighted value shown in Fig~\ref{gas}.  A temperature
gradient would be expected from a cooling flow, and the cluster around
3C~220.1 should exhibit a massive one, with the gas in the
$\beta$-model estimated to have a cooling time of $\sim 10^{10}$~yr at
a radius of 15.5 arcsec (Table~\ref{xraytab}).  However, our fits to a
straightforward cooling-flow model (Hardcastle, Worrall \&
Birkinshaw~1999) in combination with a point source are very bad, and
too spiked for the {\it Chandra\/} radial profile
(Fig.~\ref{cooling}).  The poorer angular resolution of {\it ROSAT\/}
did not provide such strong constraints, and a 30 per cent
contribution to the central emission from such a cooling-flow model
(roughly the situation shown in Fig.~\ref{cooling}) could not be ruled
out with {\it ROSAT\/}.  The temperature structure of the gas must be
complicated, so that the statistics of the current {\it Chandra\/}
data are inadequate to provide a detailed picture.

The core X-ray emission is even brighter than in our {\it ROSAT\/}
models for the source. When we use the core counts from our
two-component modelling of the {\it ROSAT\/} data, taking into account
the rather large uncertainties, and use a range of spectral parameters
within the inner contour shown in Fig.~\ref{core}, we predict $890
(+140, -500)$ {\it Chandra\/} counts in 0.3--8~keV (90 per cent
uncertainties).  Despite this large uncertainty range, the probability
of the predicted counts and the observed counts (Table~\ref{xraytab})
being the same is less than 1 per cent, suggesting variability over
the four-year interval between the {\it ROSAT\/} and {\it Chandra\/}
observations.  In contrast, a similar test for the extended component
finds agreement within errors, as expected.

Using our {\it Chandra\/} core measurement combined with data in
Table~\ref{radiotab}, we find a value for $\alpha_{\rm rx}$ (as
defined in Table~\ref{xraytab}) of 0.75.  This is less than for other
powerful radio galaxies and core-dominated quasars (see section~1),
suggesting perhaps that regions close to the AGN dominate the central
output, as is believed to be the case for lobe-dominated quasars
(Worrall et al.~1994).  The new high-resolution radio data for the
source (kindly provided by Guy Pooley; Fig.~\ref{profile}) reveal
one-sided jet emission with a jet to counter-jet ratio of $> 4$, so
that it is possible we are seeing the X-ray emission from the vicinity
of the AGN directly.  Assuming a jet speed of $0.7c$, as statistical
results for sources at comparable redshift suggest (Wardle \& Aaron
1997), the angle of the jet to the line of sight should be less than
$\sim 67$ degrees. The X-ray-measured intrinsic absorption is
relatively low (Fig~\ref{core}), consistent with expectations from the
dusty medium which HST finds to be typical in lower-power radio
galaxies like NGC~6251 (Ferrarese \& Ford 1999).  It is very different
from the case of Cygnus~A (see section~1), where the component of
X-ray emission believed to be associated with regions close to the
central black hole is seen only through a large absorbing column.
Such a low column density in 3C~220.1, coupled with X-ray emission
which is abnormally bright to be associated with the radio components,
suggests that our line of sight does not significantly intersect a
torus.  The torus may be absent, in contradiction to unified models,
but if it is present the geometry must be such as to obscure a strong
nuclear optical continuum (not evident in HST images; McCarthy et
al.~1997) while allowing us to see nuclear X-ray emission.

A major advance with {\it Chandra\/} has been the routine detection of
X-rays from kpc-scale jets, but no such emission is evident from the
eastern radio jet in 3C~220.1.  The jet in the quasar PKS~0637-752,
which is at a similar redshift to 3C~220.1, has an $\alpha_{\rm rx}$
of 0.9, based on data in Chartas et al.~(2000) for the X-ray brightest
region of knot WK7.8.  With such an $\alpha_{\rm rx}$ applied to the
9~arcsec-long radio jet of 3C~220.1 (Table~\ref{radiotab}), we
estimate 35 counts for the entire region of the jet in the {\it
Chandra\/} observation.  This corresponds to 6.6 cts arcsec$^{-2}$,
and from the radial profile in Fig.~\ref{profile} we see this would
have poor contrast against the extended thermal emission, particularly
in the inner jet regions where X-ray emission is more likely to be
seen.  The outer radio knot at 6.3~arcsec from the core contains about
2.8 mJy of the jet flux density at 8.4~GHz, and is undetected in our
X-ray data.  The corresponding $3~\sigma$ limit on the radio to X-ray
spectral index is $\alpha_{\rm rx} > 0.96$.

Similarly we might question why the radio hotspots in 3C~220.1 are not
detected with {\it Chandra\/}, but the simplest assumption of
equipartition between the magnetic-field and electron energy densities
leads to estimates of 0.15 and 0.25 counts for the eastern and western
hotspots, respectively.

Despite the relatively small sky coverage of the {\it Chandra\/}
observation, $\sim 320$ square arcmin, a second bright diffuse X-ray
source is present.  Centered off-axis (in the S2 chip), at about J2000
9~31~01.46, $+$79~13~27.8, the emission is of similar surface
brightness to 3C~220.1's cluster but covers a larger angular region.
An over-density of galaxies on deep sky survey plates suggests this
emission is associated with a less distant cluster, of redshift
between about 0.2 and 0.3.

\begin{table}
\caption{X-ray Components}
\label{xraytab}
\begin{tabular}{llc}
Parameter & Value & Notes\\
Core Counts 0.3-8 keV & $1204^{+15}_{-16}$ 
& a \\
$\alpha_x$, $N_{\rm H_{int}}$ & 0.7, $7 \times 10^{21}$~cm$^{-2}$
& b \\
Core $L_{0.7-12~\rm keV}$ & $10^{45}$ ergs s$^{-1}$ 
& b \\
Core $L_{2-10~\rm keV}$ & $6.5 \times 10^{44}$ ergs s$^{-1}$ 
& b \\
Core $S_{1~\rm keV}$ & 44 nJy 
& b \\
Core $\alpha_{\rm rx}$ & 0.75 
& h \\
$\beta$-model counts 0.3-8 keV, $\theta < 45''$ & $900^{+100}_{-120}$ &
a, c \\
$\beta$-model counts 0.3-8 keV, $\theta < 3'$ & $1400^{+550}_{-480}$ &
a, c \\
$kT$, abundance/solar & 4.6~keV, 0.16
& d \\
$\beta$-model $L_{0.7-12~\rm keV}$, $\theta < 45''$ & 5.6 $\times 10^{44}$ 
  ergs s$^{-1}$ 
& d \\
$\beta$-model $L_{2-10~\rm keV}$, $\theta < 45''$ & 3.4 $\times 10^{44}$ 
  ergs s$^{-1}$ 
& d \\
$\beta$ & $0.5^{+0.13}_{-0.065}$ 
& a, e \\
$\theta_{\rm cx}$ & $3.55^{+2.8}_{-1.5}$ 
& a, e \\
Central gas density & $(6\pm2) \times 10^{-2}$ cm$^{-3}$ 
& f \\
Gas density at $\theta = 15''$ & $6.67^{+0.23}_{-0.16} \times
10^{-3}$ cm$^{-3}$ 
& f \\
Central gas pressure & $(9.9\pm3) \times 10^{-11}$ Pa 
& f \\
Gas pressure at $\theta = 15''$ & $(1.1\pm0.03) \times 10^{-11}$ Pa 
& f \\
Central gas cooling time & $1.1^{+0.5}_{-0.3} \times 10^9$ yr 
& f \\
Cooling radius & $15.45\pm0.3$ arcsec 
& f, g \\
M$_\odot$/yr, $\theta \leq \theta_{cx}$ & $130^{+130}_{-50}$ 
& f \\
\end{tabular}
\medskip
\begin{minipage}{\linewidth}
Notes:  
a. Errors $1\sigma$ for 2 interesting parameters; 
b. Best-fit spectral parameters from Fig.~\ref{core};
c. $\beta$-model counts per unit area per unit time $\propto (1 +
  {\theta^2 \over \theta_{\rm cx}^2})^{0.5 - 3 \beta}$; 
d. Best-fit spectral parameters from Fig.~\ref{gas};
e. $\beta$ and $\theta_{\rm cx}$ highly correlated (see Fig~\ref{chi});
f. Errors $1\sigma$ for 1 interesting parameter; 
g. radius at which cooling time is 10$^{10}$ yr;
h. $\alpha_{\rm rx} = \log(l_{\rm 5~GHz}/l_{\rm 2~keV}) /7.98$.
\end{minipage}
\end{table}

\begin{table}
\caption{Radio Components}
\label{radiotab}
\begin{tabular}{ll}
Parameter & Value  \\
Core $S_{\rm 8.4~GHz}$ & $34 \pm 0.1$ mJy \\
Jet $S_{\rm 8.4~GHz}$ &  7.9 mJy  \\
Jet length & 9 arcsec \\
Jet radius & 0.3 arcsec \\
Counter Jet $S_{\rm 8.4~GHz}$ & $ < 1.95$ mJy \\
$P_{\rm jet~min}$ & 2.6 $\times 10^{-12}$ Pa \\
Western Lobe $S_{\rm 1.4~GHz}$ & 1.2 Jy \\
Lobe length & 12 arcsec \\
Lobe radius & 3.5 arcsec \\
$P_{\rm lobe~min}$ & 1.2 $\times 10^{-12}$ Pa \\
\end{tabular}
\medskip
\begin{minipage}{\linewidth}
8.4~GHz data are from map shown in Fig.~1 and lobe measurements
use the 1.4~GHz data of Harvanek \& Hardcastle (1998).
Minimum pressures assume an electron energy spectrum of number index
2.0 between $\gamma_{\rm min} = 10$ and $\gamma_{\rm max} = 10^5$
and the jet value is not corrected for possible effects of
relativistic beaming and projection (see text)
\end{minipage}
\end{table}

\section{Conclusions}

The {\it Chandra\/} data for 3C~220.1 have strikingly confirmed our
modelling of the {\it ROSAT\/} data for the source as a composite of
point-like and extended emission, and have measured the spectral
parameters of the extended and compact components with reasonable
accuracy.  The pressure in the X-ray emitting gas exceeds the minimum
pressures both in the lobes and kpc-scale jet, and the results bode
well for using powerful high-redshift radio galaxies as tracers of
galaxy groups and clusters for tests of structure-formation theories
and cosmological parameters.

There is an unsettled question pertaining to the structure of the gas
around 3C~220.1; the emissivity suggests a large cluster-scale cooling
flow, and yet the break at the boundary between the 
compact and extended emission in Fig.~\ref{profile} is too sharp
for a straightforward cooling-flow model.
This suggests that the picture of the atmosphere as slowly cooling
near a state of hydrostatic equilibrium is incorrect unless the
atmosphere is relatively young, so that a massive cooling flow has not
yet been established.  Alternatively, a central input of mechanical
energy
(from the expanding radio lobes or a merger) might be disrupting the
cooling flow.  In either case, the X-ray emission should be irregular
in appearance --- although this is not supported by the radial profile,
the statistics in the two-dimensional image (Fig.~\ref{profile}) are
not good enough to exclude such possibilities.

The X-ray core emission is bright, with an $\alpha_{\rm rx}$ closer to
that of lobe-dominated quasars than to that of core-dominated quasars,
where the core X-ray emission is dominated by a beamed component
related to the radio emission, or to that of powerful narrow-line
radio galaxies, where the core X-ray emission suffers high absorption.
This suggests that in 3C~220.1, regions close to the black hole may
dominate the central X-ray output.  Simple unification models would be
challenged if this were found to be the case for a large fraction of
high-power radio galaxies, since the models predict that the central
X-ray emission should be seen only through large absorption, with a
large deficit of photons at soft X-ray energies.  {\it Chandra\/}
observations of carefully selected samples of high-power radio
galaxies can test these models.

\section*{Acknowledgments}

We are grateful to Guy Pooley for kindly providing the 8.4~GHz VLA A
and B array data for 3C~220.1.  We thank staff of the CXC for help
concerning calibrations and the {\sc ciao} and {\sc ds9} software.

\label{lastpage}

\end{document}